\documentstyle[11pt]{article}
\begin{document}
\newcommand{\be}{\begin{equation}}
\newcommand{\ee}{\end{equation}}
\newcommand{\bea}{\begin{eqnarray}}
\newcommand{\eea}{\end{eqnarray}}
\newcommand{\beaa}{\begin{eqnarray*}}
\newcommand{\eeaa}{\end{eqnarray*}}
\newcommand{\qd}{\quad}
\newcommand{\qqd}{\qquad}
\newcommand{\npb}{\nopagebreak[1]}
\newcommand{\nn}{\nonumber}
\newcommand{\prel}{\Preliminaries}
\newcommand{\theor}{Theorem}
\newtheorem{defi}{Definition}
\newtheorem{prop}{Proposition}
\newtheorem{rem}{Remark}
\newcommand{\example}{Example}
\title{ \bf On the Law of Transformation of Affine
Connection and its Integration. \\
Part 1. Generalization of the Lame equations}
\author{V.S. Dryuma\thanks{Work supported  by MURST,Italy.
Permanent address: Institute of Mathematics
 Academy of Sciences of Moldova
Kishinev, MD 2028, Moldova; Academitcheskaya str.5, e-mail:
15valery@mathem.moldova.su, valery@gala.moldova.su}}
%\\ \em Institute of Mathematics Moldavian Academy of sciences,\\ \em
%\\ 277028, Academitcheskaya str.5, Kishinev, Moldova}
\date{}
\maketitle
\begin{abstract}
The law of transformation of affine connection for n-dimensional
manifolds as the system of nonlinear equations on local coordinates of manifold
is considered. The extension of the Darboux-Lame system of equations to
the spaces of constant negative curvature is demonstrated.
Geodesic deviation equation as well as the equations of geodesics
are presented in the form of the matrix Darboux-Lame system of equations.
\end{abstract}
%\end{document}
%\newpage

\section{Introduction}

     It is well-known that the integrable equations have numerous
applications in geometry. The Korteveg-de Vries equation, sin-Gordon equation, Tzitzeika
equation, Ka\-dom\-tzev-Petviashvili equation, Zakharov-Manakov system of
equations and others are the most famous examples of such type of equations.
This can be explained by the fact that  the mentioned above equations
have the Lax pair representation, which is equivalent to the condition
of zero curvature for suitable connections.
From this we infer that the law of transformation of affine connection
is a key to understanding the nature of such type of equations and their
integrability.

     The matrix Zakharov-Manakov system of equations discovered in context of
formal generalization of Inverse Scattering Transform Method in
multi-dimensional case [1] after its geometrical interpretation as the
Darboux-Lame system [2--6] gives an example of such relation between equations and
geometry because it corresponds to the simplest case of the law of
transformation of the flat affine connection.
 For the  spaces of non-zero curvature  the law of
transformation of connection leads to generalization of Zakharov-Manakov
and  Darboux-Lame systems of equations.

     The problem of integration of equations of geodesics has a great
significance in geometry. We show here that in a certain case this problem is
connected with the matrix Darboux-Lame system  of equations
and its generalizations.

     The equation of geodesic deviation is also important in geometry,
in the theory of Riemann spaces. For surfaces it coincides, in essence, with
the Gauss equation in
geodesic coordinates while in the three- and higher-dimensional Riemann spaces
it carries essential information about these spaces. Solutions of the
geodesic deviations equation (Jacoby fields) and their properties (e.g., the
existence of conjugate points) are related to various important
characteristics of Riemann spaces.

     In this paper we also consider applications of the
 matrix Darboux-Lame system to the study of the geodesic deviation equation.

\section{ Equation of the law of transformation of affine connection}

     Here we will present for convenience basic formulas, which will be used
in what follows.
Let $M^n$ be the  space  endowed with the affine connection. This means that
the components of connections $\Gamma^i_{jk}(x^l)$ and $\bar\Gamma^i_{jk}(y^l)$
in two various systems of coordinates $(x^l)$ and $(y^l)$ are connected by the
relations
\begin{equation}
\frac{\partial^2 y^k}{\partial x^i \partial x^j}=\Gamma^l_{ij}(x^s)\frac
{\partial y^k}{\partial x^l}-\bar \Gamma^k_{lm}(y^s)\frac
{\partial y^l}{\partial x^i} \frac{\partial y^m}{\partial x^j}.
 \label{law}
\end{equation}

      We can consider these relations as the system of partial differential
equations for $n$ unknown functions $y^k$ of variables $x^k$. In general case
this is a compatible system of equations and its properties depend on the
curvature tensor
\be
R^i_{kln}=\left[\frac{\partial\Gamma_n}{\partial x^l}-
\frac{\partial\Gamma_l}{\partial x^n}+\Gamma_l\Gamma_n -
\Gamma_n\Gamma_l\right]^i_k,
\label{ten}
\ee

which has the law of transformation as the four rank tensor
$$
R^i_{kln}=\frac{\partial x^i}{\partial y^s}\frac{\partial y^r}{\partial x^k}
\frac{\partial y^q}{\partial x^l}\frac{\partial y^p}{\partial x^n}
\bar R^s_{rqp},
$$
where $\Gamma_l$ are  matrices with components $\Gamma^i_{lk}$ and
$R^i_{kln}$ are  components of curvature tensor of the connection
$\Gamma_l$.

     The system ~(\ref{law}) for manifolds of dimension three
is the main object of our consideration.

\section{ Triply-orthogonal systems of surfaces, flat coordinates and the
Darboux-Lame system of equations}

     The above system of equations  in three-dimensional case is the
system of 18 equations for three unknown functions $u(x,y,z)$, $v(x,y,z)$,
$w(x,y,z)$, depending on three variables $(x,y,z)$. It contains 18
coefficients of affine connection of the manifold and their solutions
depend on
the specific choice of coefficients of connection. With the help of
coordinates $u(x,y,z)$, $v(x,y,z)$, $w(x,y,z)$ the geometry  of the 3-dim
space is described. Let us consider the simplest  of examples.

     It corresponds to the flat space, i.e. when the coefficients
of connection  $\bar\Gamma^i_{jk}\equiv 0$, or the tensor of curvature
of the connection $\Gamma^i_{jk}$  is equal to zero:

      The initial system ~(\ref{law}) takes the form
\be
\frac{\partial^2 y^k}{\partial x^i \partial x^j}=\Gamma^l_{ij}(x^s)\frac
{\partial y^k}{\partial x^l}.\label{lam}
\ee

     For 3-dim space we have the following system of equations
$$
\frac{\partial^2 u}{\partial x \partial y}= \Gamma^1_{12}\frac{\partial u}
{\partial x}+\Gamma^2_{12}\frac{\partial u}{\partial y}+\Gamma^3_{12}
\frac{\partial u}{\partial z},\quad
\frac{\partial^2 u}{\partial x \partial z}= \Gamma^1_{13}\frac{\partial u}
{\partial x}+\Gamma^2_{13}\frac{\partial u}{\partial y}+\Gamma^3_{13}
\frac{\partial u}{\partial z},
$$
$$
\frac{\partial^2 u}{\partial y \partial z}= \Gamma^1_{23}\frac{\partial u}
{\partial x}+\Gamma^2_{23}\frac{\partial u}{\partial y}+\Gamma^3_{23}
\frac{\partial u}{\partial z},\quad
\frac{\partial^2 u}{\partial x^2}= \Gamma^1_{11}\frac{\partial u}
{\partial x}+\Gamma^2_{11}\frac{\partial u}{\partial y}+\Gamma^3_{11}
\frac{\partial u}{\partial z},
$$
$$
\frac{\partial^2 u}{\partial y^2}= \Gamma^1_{22}\frac{\partial u}
{\partial x}+\Gamma^2_{22}\frac{\partial u}{\partial y}+\Gamma^3_{22}
\frac{\partial u}{\partial z},\quad
\frac{\partial^2 u}{\partial z^2}= \Gamma^1_{33}\frac{\partial u}
{\partial x}+\Gamma^2_{33}\frac{\partial u}{\partial y}+\Gamma^3_{33}
\frac{\partial u}{\partial z},
$$
and corresponding equations for the coordinates $v(x,y,z)$ and
$w(x,y,z)$.

     The system of equations ~(\ref{lam}) is very well known in
classical Differential Geometry. This is the Lame system of equations for the
triply orthogonal curvilinear  coordinate systems in a flat Euclidean 3-dim
space. At present the half of this system containing only mixed
derivatives has appeared in an non-evident general matrix form
in work [1] in context of the  multidimensional generalization of
integrable differential equations but without any applications.
An explicit form of this system of equations with its  clear geometrical
sense, simplest solutions and applications was presented by the author
(see [2-6]).

     Here I briefly review some results of this approach.
For the Riemann 3-dim space in orthogonal metric
$$
d{s}^{2} ={A}^{2}(x,y,z)d{x}^{2}+{B}^{2}(x,y,z)d{y}^{2}+
{C}^{2}(x,y,z)d{z}^{2}
$$
the condition
$$
R_{ijkl} ={\kappa}(g_{ik}g_{jl}-g_{il}g_{jk})
$$
leads to six equations, three equations of which do not contain parameter
$\kappa$.

\begin{eqnarray}
&A_{zy}=(C_y/C)A_z+(B_z/B)A_y, \nonumber \\
&B_{zx}=(C_x/C)B_z+(A_z/A)B_x,\nonumber \\
&C_{xy}=(A_y/A)C_x+(B_x/B)C_y. \nonumber
\end{eqnarray}

     These equations named the Darboux system can be considered as the
scalar reduction of the general
matrix Zakharov-Manakov system of equations [1] and they can be integrated by
the Inverse Scattering Method using the following linear problem:

\begin{eqnarray}
&{\Phi}_{zy}=(C_y/C){\Phi}_z+(B_z/B){\Phi}_y,\nonumber \\
&{\Phi}_{zx}=(C_x/C){\Phi}_z+(A_z/A){\Phi}_x,\nonumber \\
&{\Phi}_{xy}=(A_y/A){\Phi}_x+(B_x/B){\Phi}_y.\nonumber 
\end{eqnarray}

     The partial case of the Darboux system of equations is connected 
with theory of the normal Riemann space. The notion of normal Riemann space 
was introduced by Eisenhart.
\begin{defi} 

     The n-dimensional Riemmanian space with local coordinates $u^i$
is normal when the conditions on main curvatures $K_i$ is fulfiled 
$$
\frac{\partial K_l}{\partial u^l}=3\lambda_l+3 \mu_l K_l,
$$
and
$$
\frac{\partial K_i}{\partial u^l}=\lambda_l+ \mu_l K_i,\quad i \not=l,
$$
$$
\frac{\partial \ln g_{ij}}{\partial u^l}=\frac{2}{K_l-K_i}
\frac{\partial K_i}{\partial u^l} \quad i \not=l,
$$
where $\lambda$ and $\mu$ are some functions of coordinates $u^i$.     
\end{defi}

\begin{rem}
      The values $K_1, K_2,\cdots,K_n$  are given the name of the 
principial curvatures relatively to some symmetrical tensor $b_{ij}$ of n-dimensional 
Riemmanian space $M^n$ with metric
$$
ds^2=g_{ij}du^idu^j
$$
if they are the roots of algebraic equation
$$
|b_{ij}-K g_{ij}|=0.
$$
  
     The inherent vectors $\xi^i_h$ of main directions of the tensor $b_{ij}$ 
from the equations
$$
(b_{ij}-K_h g_{ij})\xi^i_h=0
$$
are defined. They are orthogonal
$$
g_{ij}\xi^i_p \xi^j_q=0 \quad p\not=q,
$$
and satisfy to the conditions
$$
b_{ij}\xi^i_p \xi^j_q=0, \quad p\not=q.    
$$

     According to [7] the system of equations for the principal curvatures         
in the 3-dim case looks  as
$$
(K_2-K_3)K_{1x}+3(K_3-K_1)K_{2x}+3(K_1-K_2)K_{3x}=0,
$$
$$
3(K_2-K_3)K_{1y}+(K_3-K_1)K_{2y}+3(K_1-K_2)K_{3y}=0,
$$
$$
3(K_2-K_3)K_{1z}+3(K_3-K_1)K_{2z}+(K_1-K_2)K_{3x}=0.
$$

    Using the relations
\begin{eqnarray}
\frac{A_y}{A}=\frac{K_{1y}}{K_2-K_1}, \quad
\frac{A_z}{A}=\frac{K_{1z}}{K_3-K_1},\quad
\frac{B_x}{B}=\frac{K_{2x}}{K_1-K_2},\nonumber \\
\frac{B_z}{B}=\frac{K_{2z}}{K_3-K_2},\quad
\frac{C_x}{C}=\frac{K_{3x}}{K_1-K_3},\quad 
\frac{C_y}{C}=\frac{K_{3y}}{K_2-K_3},\nonumber
\end{eqnarray}
we get
$$
K_{1xy}+\frac{A_y}{A}K_{1x}+\left( \frac{B_x}{B}+\frac{A_x}{A}-\frac{A_{xy}}
{A_y}\right)K_{1y}=0,    
$$
$$
K_{1xz}+\frac{A_z}{A}K_{1x}+\left( \frac{C_x}{C}+\frac{A_x}{A}-\frac{A_{xz}}
{A_z}\right)K_{1z}=0    
$$
$$
K_{1zy}+ \left (\frac{A_z}{A}-\frac{C_y A_z}{C A_y}\right ) K_{1y}+ \left ( \frac{C_y}{C}+
\frac{A_y}{A}-\frac{A_{zy}}{A_z}\right)K_{1z}=0,    
$$
or
$$
K_{1zy}+\left(\frac{A_y}{A}-\frac{B_z A_y}{B A_z}\right ) K_{1z}+\left( \frac{B_z}{B}+
\frac{A_z}{A}-\frac{A_{zy}}{A_z}\right)K_{1y}=0.    
$$

     By analogic way the system of equations for $K_2$ and $K_3$ have been
written.
 
     For  the construction of partial solutions of this system we can use
the linear system of equations
\begin{eqnarray}
&{\Phi}_{zy}+\frac{1}{2(z-y)})\left[{\Phi}_z-{\Phi}_y\right]=0,\nonumber \\
&{\Phi}_{zx}+\frac{1}{2(z-x)}\left[{\Phi}_z-{\Phi}_x\right]=0,\label{dar} \\
&{\Phi}_{xy}+\frac{1}{2(x-y)}\left[{\Phi}_x-{\Phi}_y\right]=0.\nonumber
\end{eqnarray}

     If $\varphi$ and $\chi$ are two solutions of the system ~(\ref{dar})
connected by the relation:
\be
\frac{\partial \varphi}{\partial u_i}=\chi/2-u_i\frac{\partial
 \chi}{\partial u_i}, \label{sob}
\ee
where $u_i =(x,y,z)$,
then the equations ~(\ref{dar}) are the conditions of compatibility
for ~(\ref{sob}).

      So, if the function $\chi_h$ is a solution of the system 
~(\ref{dar}),
then the function $\varphi_h$ also will be a solution. As result we have the
following relations between the solutions
$$
\frac{\partial \chi_{h+1}}{\partial u_i}=
\chi_h/2-u_i\frac{\partial \chi_h}{\partial u_i},
$$

     As example, begining from the trivial solution $\chi_0=0$ we can
obtain the solutions in the form of symmetrical functions of variables
$(x,y,z)$ [5, 7].
\end{rem}

     Recently the Darboux and the full Lame system of equations have been
integrated by the IST-method [8,9,10].

\begin{rem}
The full Lame system of equations can be applied to integration of the Einstein
equations. The simplest illustration of that is the result of
E.Kasner on the representation of the Schwarzschild solution of the Einstein equations
$$
ds^2=(1-\frac{2m}{r})dt^2-(1-\frac{2m}{r})^{-1}dr^2-r^2d \theta^2-
r^2\sin^2 \theta d \varphi^2
$$
as six squares of differentials in 6-dim flat space
$$
ds^2=-dx^2-dy^2-dz^2+dX^2+dY^2-dZ^2,
$$
where X, Y, Z are defined by equations:
$$
Z=\int \sqrt {1+\frac{256 m^4}{(R^2+16 m^2)^3}}dR,
\quad X=\frac{R \sin t}{\sqrt { R^2+16 m^2}},
$$
$$
Y=\frac{R \cos t}{\sqrt {R^2+16 m^2}},
\quad R=\sqrt {8 m(r-2 m)}.
$$

    Using this result we can obtain the Schwarzschild solutions of the Einstein
equations from the Lame system for the flat six-dimensional space.

    It is apparent that  arbitrary 4-dim metric can be presented as
embedded in a flat space of suitable dimension and use then the theory of
the Lame
equations for integration  of the Einstein equations.
\end{rem}

\begin{rem}
     The theory of the Lame system of equations is connected with the Euler-Picard
system of partial differential equations, having important  applications to
number theory. The equations have the form
$$
\frac{\partial^2 F}{\partial t_i \partial t_j}+\frac{l}{m(t_j-t_i)}
\left (\frac{\partial F}{\partial t_i}-\frac{\partial F}{\partial t_j}\right )=0,
$$
$$
\frac{\partial^2 F}{\partial t_k^2}+\frac{l}{n (t_k-1)t_k}\Bigl\{\sum^r_{{i=1},
{i\not=k}}\frac{t_i^2-t_i}{t_i-t_k}\frac{\partial }{\partial t_i}-
\Bigl (\sum^r_{{i=1},{i\not=k}}\frac{(t_k-1)t_i}{t_i-t_k}-
$$
$$
- (r+2)(t_k-1)-t_k-1) \Bigr)\frac{\partial }{\partial t_k}+
\frac{l(r+2)-n}{n} \Bigr \} F=0,
$$
where $l$ and $n$ are natural numbers such that $0 < l < n$
and the integral functions
$$
\int_{\alpha_t}\frac{dx}{y^l}
$$
of variables $t=t_1...t_r$ and of the family of cycles $\alpha_t$ on compact
Riemann surfaces of planar equations of the type
$$
Y^n=(X-1)X(X-t_1)(X-t_2)...(X-t_r),
$$
are solutions of the Euler-Picard system [11].
\end{rem}

\section{ On equations for the coordinates of distorted spaces}

    The simplest generalization of the Lame  system of equations is connected
with the manifolds of constant negative curvature.  They have the metric
$$
ds^2=\frac{dx^2 +dy^2 +dz^2}{z^2}
$$
and for such a metric the matrices of the Christoffel's symbols are
$$
\bar\Gamma_1=\left | \begin{array}{ccc}
0 & 0 & - \frac{1}{w} \\
0 & 0 & 0 \\
\frac{1}{w} & 0 & 0 
\end{array} \right|, \
\bar\Gamma_2=\left\Vert \begin{array}{ccc}
0 & 0 & 0 \\
0 & 0 & -\frac{1}{w} \\
0 & \frac{1}{w} & 0
\end{array} \right |, \
\bar\Gamma_3=\left| \begin{array}{ccc}
-\frac{1}{w} & 0 & 0 \\
0 & -\frac{1}{w} & 0 \\
0 & 0 & -\frac{1}{w}
\end{array} \right |.
$$
     From the fundamental system of equations (1) for the coordinates
we get
$$
u_{xy}= (\Gamma^1_{12}+\frac{w_y}{w}) u_x+(\Gamma^2_{12}+\frac{w_x}{w}) u_y,
$$
$$
u_{xz}= (\Gamma^1_{13}+\frac{w_z}{w}) u_x+(\Gamma^3_{13}+\frac{w_x}{w}) u_z,
$$
$$
u_{yz}= (\Gamma^2_{23} +\frac{w_z}{w}) u_y+(\Gamma^3_{23}+\frac{w_y}{w}) u_z,
$$
$$
v_{xy}= (\Gamma^1_{12}+\frac{w_y}{w}) v_x+(\Gamma^2_{12}+\frac{w_x}{w}) v_y,
$$
$$
v_{xz}= (\Gamma^1_{13}+\frac{w_z}{w}) v_x+(\Gamma^3_{13}+\frac{w_x}{w}) v_z,
$$
$$
v_{yz}= (\Gamma^2_{23} +\frac{w_z}{w}) v_y+(\Gamma^3_{23}+\frac{w_y}{w}) v_z,
$$
where the coefficients $\Gamma^3_{12}$, $\Gamma^2_{12}$, and $\Gamma^1_{12}$
can be reduced to zero using the special choice of system of coordinates.

     The first three of above equations have the form of the Darboux system
of equations and the equations for the coordinate $w(x,y,z)$ can be written
as follows:
$$
w_{xy}+\frac{w_x w_y}{w}= \frac{A_y}{A} w_x+\frac{B_x}{B} w_y-\frac
{u_x u_y+v_x v_y}{w},
$$
$$
w_{xz}+\frac{w_x w_z}{w}= \frac{A_z}{A} w_x+\frac{C_x}{C} w_z-\frac
{u_x u_z+v_x v_z}{w},
$$
$$
w_{yz}+\frac{w_y w_z}{w}= \frac{B_z}{B} w_y+\frac{C_y}{C} w_z-\frac
{u_y u_z+v_y v_z}{w}.
$$

     After the change of variable $w=\sqrt R$ we obtain the following
compatible system of equations
$$
R_{zy}=\frac{B_z}{B} R_y+\frac{C_y}{C} R_z-2(u_y u_z+v_y v_z),
$$
\be
R_{zx}= \frac{A_z}{A} R_x+\frac{C_x}{C} R_z-2(u_x u_z+v_x v_z),\label{lob}
\ee
$$
R_{xy}=\frac{A_y}{A}R_x+\frac{B_x}{B}R_y-2(u_x u_y+v_x v_y).
$$

     The solutions of this system of equations can be used to study
problems of the theory of 3-dimensional manifolds, the knots theory and so on.

     It is apparent that in an anologous way one can  investigate the
Einstein spaces and other more general affine-connected spaces.

\begin{rem}
      It can be shown that the fundamental system of equations ~(\ref{law})
 (and ~(\ref{lob})
can be integrated (using some modification!) with the help of representation
of the functions $u^i$ in the form
$$
u^i(x,y,z)=U^i(x,y,z)\exp[\frac{x}{\lambda}+\frac{y}{\lambda-1}+
\frac{z}{\lambda+1}].
$$

      Corresponding solutions for connections coefficients and coordinates
$u^i$ of space have important applications to different problems of Geometry.
\end{rem}

\section{ On geodesics  of the space of affine connection}

     The problem of integration of equations of geodesics is
very important in  Differential Geometry.
The simplest case of such type of equations is connected with the second
order ODE
$$
y''+a_1(x,y){y'}^3+ 3a_2(x,y){y'}^2+3a_3(x,y)y'+a_4(x,y)=0.
$$
It turns out that the theory of the matrix Laplace equation is connected with
this problem.

    The general equations of geodesics of the affine connected spaces
with coefficients $\Gamma^k_{ij}$ are:
$$
\frac{d^2 x^i}{ds^2}+\Gamma^i_{kj}\frac{d x^k}{ds}\frac{d x^j}{ds}=0.
$$

     Let us  present them in the following form
$$
\frac{d^2 x}{ds^2}+\Gamma^1_{11}\left(\frac{d x}{ds}\right)^2+
2\Gamma^1_{12}\frac{d x}{ds}\frac{dy}{ds}+
\Gamma^1_{22}\left(\frac{d y}{ds}\right)^2+
2\Gamma^1_{1j}\frac{d x}{ds}\frac{d z^j}{ds}+
$$
$$
+2\Gamma^1_{2j}\frac{d y}{ds}\frac{d z^j}{ds}+
\Gamma^1_{kj}\frac{d z^k}{ds}\frac{d z^j}{ds}=0,
$$
$$
\frac{d^2 y}{ds^2}+\Gamma^2_{11}\left(\frac{d x}{ds}\right)^2+
2\Gamma^2_{12}\frac{d x}{ds}\frac{dy}{ds}+
\Gamma^2_{22}\left(\frac{d y}{ds}\right)^2+
2\Gamma^2_{1j}\frac{d x}{ds}\frac{d z^j}{ds}+
$$
$$
+2\Gamma^2_{2j}\frac{d y}{ds}\frac{d z^j}{ds}+
\Gamma^2_{kj}\frac{d z^k}{ds}\frac{d z^j}{ds}=0,
$$
$$
\frac{d^2 z^i}{ds^2}+\Gamma^i_{11}\left(\frac{d x}{ds}\right)^2+
2\Gamma^i_{12}\frac{d x}{ds}\frac{dy}{ds}+
\Gamma^i_{22}\left(\frac{d y}{ds}\right)^2+
2\Gamma^i_{1j}\frac{d x}{ds}\frac{d z^j}{ds}+
$$
$$
+2\Gamma^i_{2j}\frac{d y}{ds}\frac{d z^j}{ds}+
\Gamma^i_{kj}\frac{d z^k}{ds}\frac{d z^j}{ds}=0,
$$
writing the  coordinates $x^i$ in the form  $x^i=(x,y,z^i)$.

      Then we will consider the coordinates $z^i$ as the functions of
variables  $x,y$. So the following relations are fulfilled:
$$
\frac{dz^i}{ds}= z^i_{x}\frac{dx}{ds}+z^i_{y} \frac{dy}{ds}
$$
and
$$
\frac{d^2 z^i}{ds^2}= z^i_{xx}\left(\frac{dx}{ds}\right)^2+
2z^i_{xy}\frac{dx}{ds}\frac{dy}{ds} + z^i_{yy}\left(\frac{dy}{ds}\right)^2
+ z^i_x \frac{d^2 x}{ds^2}+ z^i_y \frac{d^2 y}{ds^2}.
$$

     Putting these relations in above formulas one obtains
the system of equations for the functions $z^i(x,y)$ (after making equal to
zero the expressions at the derivatives $(\frac{d x}{ds})^2,\
(\frac{dy}{ds})^2 $ and $ (\frac{d x}{ds}\frac{dy}{ds})$.
$$
z^i_{xx}=[\Gamma^1_{11}+2\Gamma^1_{1j}z^j_x+\Gamma^1_{kj}z^k_x z^j_x] z^i_x+
[\Gamma^2_{11}+2\Gamma^2_{1j}z^j_x+\Gamma^2_{kj}z^k_x z^j_x] z^i_y -
2\Gamma^i_{1j}z^j_x-\Gamma^i_{11}-\Gamma^i_{kj}z^k_xz^j_x,
$$
$$
z^i_{xy}=[\Gamma^1_{12}+\Gamma^1_{1j}z^j_y+\Gamma^1_{2j}z^j_x+
\frac{1}{2}\Gamma^1_{kj}(z^k_x z^j_y+z^k_y z^j_x)] z^i_x +
$$
$$
+[\Gamma^2_{12}+\Gamma^2_{1j}z^j_y+\Gamma^2_{2j} z^j_x+
\frac{1}{2}\Gamma^2_{kj}(z^k_x z^j_y+z^k_y z^j_x)] z^i_y-
$$
$$
-\Gamma^i_{1j}z^j_y - \Gamma^i_{2j} z^j_x-
\Gamma^i_{12}-\frac{1}{2}\Gamma^i_{kj}
(z^k_x z^j_y+z^k_y z^j_x),
$$
$$
z^i_{yy}=[\Gamma^1_{22}+2\Gamma^1_{2j} z^j_y+\Gamma^1_{kj} z^k_y z^j_y] z^i_x+
[\Gamma^2_{22}+2\Gamma^2_{2j} z^j_y+\Gamma^2_{kj} z^k_y z^j_y] z^i_y -
2\Gamma^i_{2j}z^j_y-\Gamma^i_{22}-\Gamma^i_{kj}z^k_y z^j_y.
$$

     So one obtains the following statement

\begin{prop}
     There is one-to-one correspondence between the second order ODE
$$
y''+a_1(x,y){y'}^3+ 3a_2(x,y){y'}^2+3a_3(x,y)y'+a_4(x,y)=0.
$$
and two-dimensional surfaces of the affine connected space $A^n$.

     The following relations between the coefficients $a_i(x,y)$ of the
equation, the coordinates  $z^k(x,y)$ of the space $A^n$
and the coefficients of the  connections $\Gamma^k_{ij}(x,y)$ are true
$$
y''-[\Gamma^1_{22}+2\Gamma^1_{2j}z^j_y+\Gamma^1_{kj} z^k_y z^j_y]\ {y'}^3+
[\Gamma^2_{22}-2\Gamma^1_{12}+\Gamma^2_{kj}z^k_y z^j_y-2\Gamma^1_{2j}z^j_x -
2\Gamma^1_{1j}z^j_y+
$$
$$
+2\Gamma^2_{2j}z^j_y-
\Gamma^1_{kj}(z^k_xz^j_x+z^j_x z^k_y)]\ {y'}^2+
$$
$$
+[2\Gamma^2_{12}-\Gamma^1_{11}+ \Gamma^2_{kj}(z^k_x z^j_y+z^j_x z^k_y)+
2\Gamma^2_{2j}z^j_x+2\Gamma^2_{1j}z^j_y-2\Gamma^1_{1j}z^j_x-
\Gamma^1_{kj}z^k_x z^j_x]\ y'+
$$
$$
+\Gamma^2_{11}+2\Gamma^2_{1j}z^j_x+\Gamma^2_{kj}z^k_x z^j_x=0.
$$
\end{prop}

      The above mentioned system of equations for coordinates $z^i(x,y)$
in general case is nonlinear generalization of the matrix Laplace
equations.

      Restricting our consideration to the linear part of this system of
equations we get
\begin{eqnarray}
&Z_{xx}= A(x,y) Z_x + B(x,y) Z_y, \nonumber \\
&Z_{xy}= C(x,y) Z_x + D(x,y) Z_y, \\
&Z_{yy}= E(x,y) Z_x + F(x,y) Z_y,\nonumber
\end{eqnarray}
where $\vec Z=z^i(x,y)$ is a vector-function, $A,B,C,D,E,F$ are matrices.

From the conditions of compatibility the relations follow:
$$
A_y-C_x=[C,A]+DC-BE,
$$
\be
D_y-F_x=[FD]+EB-CD,
\ee
$$
D_x-B_y=AD+BF-CB-D^2,
$$
$$
C_y-E_x=EA+FC-DE-C^2.
$$
    Let us consider some particular cases.

     1. $N=3$. The linear system is:
$$
Z_{xx}=(\Gamma^1_{11}-2\Gamma^3_{13})Z_x +\Gamma^2_{11} Z_y,
$$
$$
Z_{xy}=(\Gamma^1_{12}-\Gamma^3_{23})Z_x +(\Gamma^2_{12}-\Gamma^3_{13}) Z_y,
$$
$$
Z_{yy}=\Gamma^1_{22} Z_x +(\Gamma^2_{22}-2\Gamma^3_{23}) Z_y.
$$
From the condition of compatibility $Z_{xxy}=Z_{xyx}$ and $Z_{yyx}=Z_{xyy}$
we get:
$$
\Gamma^3_{13}=\frac{1}{3}(\Gamma^1_{11}+\Gamma^2_{12}),\quad
\Gamma^3_{23}=\frac{1}{3}(\Gamma^1_{12}+\Gamma^2_{22}),\quad
\Gamma^2_{23}=\Gamma^1_{13}
$$
and as result we obtain the system of equations
$$
Z_{xx}=\frac{1}{3}(\Gamma^1_{11}-2\Gamma^2_{12})Z_x +\Gamma^2_{11} Z_y,
$$
$$
Z_{xy}=\frac{1}{3}(2\Gamma^1_{12}-\Gamma^2_{22})Z_x +
\frac{1}{3}(2\Gamma^2_{12}-\Gamma^1_{11}) Z_y,
$$
$$
Z_{yy}=\Gamma^1_{22} Z_x +\frac{1}{3}(\Gamma^2_{22}-2\Gamma^1_{12})Z_y.
$$

     The corresponding equations of geodesics are
\be
y''-\Gamma^1_{22}{y'}^3+ (\Gamma^2_{22}-2\Gamma^1_{12}){y'}^2+
(2\Gamma^2_{12}-\Gamma^1_{11})y'+\Gamma^2_{11}=0.\label{geo}
\ee

     From the conditions of compatibility one obtains the equations
$$
a_{1x}-a_{2y}=2a_3a_1-2a_{2}^2,\quad
a_{2x}-a_{3y}=a_1a_4-a_2 a_3,
$$
$$
a_{4y}-a_{3x}=-2a_2 a_4+2a_{3}^2.
$$

     These conditions for coefficients $a_i$  correspond to the ODE
(9) with the projective flat of connection, i.e. the components
of its curvature tensor are equal to zero [12,13].

      2. N=4
In linear approximation we obtain
$$
z^3_{xx}=(\Gamma^1_{11}-2\Gamma^3_{13})z^3_x+\Gamma^2_{11}z^3_y-
2\Gamma^3_{14}z^4_x,
$$
$$
z^3_{xy}=(\Gamma^1_{12}-\Gamma^3_{23})z^3_x+(\Gamma^2_{12}-
\Gamma^3_{13})z^3_y -\Gamma^3_{24}z^4_x-\Gamma^3_{14}z^4_y,
$$
$$
z^3_{yy}=\Gamma^1_{22}z^3_x+(\Gamma^2_{22}-2\Gamma^3_{23})z^3_y-
2\Gamma^3_{24}z^4_y,
$$
$$
z^4_{xx}=(\Gamma^1_{11}-2\Gamma^4_{14})z^4_x+\Gamma^2_{11}z^4_y-
2\Gamma^4_{13}z^3_x,
$$
$$
z^4_{xy}=(\Gamma^1_{12}-\Gamma^4_{24})z^4_x+(\Gamma^2_{12}-
\Gamma^4_{14})z^4_y -\Gamma^4_{13}z^3_y-\Gamma^4_{23}z^3_x,
$$
$$
z^4_{yy}=\Gamma^1_{22}z^4_x+(\Gamma^2_{22}-2\Gamma^4_{24})z^4_y-
2\Gamma^4_{23}z^3_y.
$$

     Or in the matrix form
$$
Z_{xx}=\left ( \begin{array}{cc}
\Gamma^1_{11}-2\Gamma^3_{13} & -2\Gamma^3_{14} \\ 
-2\Gamma^4_{13} & \Gamma^1_{11}-2\Gamma^4_{14}
\end{array} \right) Z_x + \left( \begin{array}{cc}
\Gamma^2_{11} & 0 \\
0 & \Gamma^2_{11}
\end{array} \right) Z_y,
$$
$$
Z_{xy}=\left( \begin{array}{cc}
\Gamma^1_{12}-\Gamma^3_{23} & -\Gamma^3_{24}\\
-\Gamma^4_{23} & \Gamma^1_{12}-\Gamma^4_{24}
\end{array} \right) Z_x + \left( \begin{array}{cc}
\Gamma^2_{12}-\Gamma^3_{13} & -\Gamma^3_{14}\\
-\Gamma^4_{13} & \Gamma^2_{12}-\Gamma^4_{14}
\end{array} \right) Z_y,
$$
$$
Z_{yy}=\left( \begin{array}{cc}
\Gamma^1_{22} & 0\\
0 & \Gamma^1_{22}
\end{array} \right) Z_x + \left( \begin{array}{cc}
\Gamma^2_{22}-2\Gamma^3_{23} & -2\Gamma^3_{24}\\
-2\Gamma^4_{23} & \Gamma^2_{22}-2\Gamma^4_{24}
\end{array} \right) Z_y.
$$

     Let us consider the surfaces in  4-dim space corresponding to
the equations
$y''= f(x,y).$  According to (9) one get
$$
y''+\Gamma^2_{11}=0,
$$
and the following relations are true:
$$
\Gamma^1_{22}=0,\quad \Gamma^2_{22}-2\Gamma^1_{12}=0,\quad
2\Gamma^2_{12}-\Gamma^1_{11}=0,\quad \Gamma^2_{11}\not=0.
$$

     From these relations  we get
$$
F=2C,\quad A=2D,\quad E=0,
$$
and the system  takes the form
$$
C_x-2D_y+2CD-DC=0,\quad
D_x-B_y+CB-2BC-D^2=0,
$$
$$
D_y-2C_x+2DC-CD=0,\quad
C_y=C^2.
$$

      So from the first and third equations we have
$$
C_x-D_y+CD-DC=0,
$$
and the matrices $C$ and  $D$ can be presented in the form
$$
 C=\Theta_y\Theta^{-1},\quad D=\Theta_x\Theta^{-1},
$$
where $\Theta(x,y)$ is matrix satisfying  the compatible system of
equations
$$
\Theta_{xy}=\Theta_x\Theta^{-1}\Theta_{y}+\Theta_y \Theta^{-1}\Theta_x
$$
$$
\Theta_{yy}=2\Theta_{y}\Theta^{-1}\Theta_{y}.
$$

     The foregoing method of investigation of equations of geodesics
can be generalized for the spaces of higher dimensions.
So in three-dimensional case we obtain
the following equation of geodesics
$$
y''-\Gamma^1_{22}{y'}^3+ (\Gamma^2_{22}-2\Gamma^1_{12}){y'}^2+
(2\Gamma^2_{12}-\Gamma^1_{11})y'+\Gamma^2_{11}+
$$
$$
+2\Gamma^2_{13}z'+\Gamma^2_{33}{z'}^2+2(\Gamma^2_{23}-\Gamma^1_{13})y'z'-
2\Gamma^1_{23}{y'}^2z'-\Gamma^1_{33}y'{z'}^2=0,
$$
$$
z''-\Gamma^1_{33}{z'}^3+ (\Gamma^3_{33}-2\Gamma^1_{13}){z'}^2+
(2\Gamma^3_{13}-\Gamma^1_{11})z'+\Gamma^3_{11}+
$$
$$
+2\Gamma^3_{12}y'+\Gamma^3_{22}{y'}^2+2(\Gamma^3_{23}-\Gamma^1_{12})y'z'-
2\Gamma^1_{23}y'{z'}^2-\Gamma^1_{22}{y'}^2z'=0.
$$

     From these relations the matrix
Darboux-Lame system has appeared by natural way.

     In fact, the coordinates $x^i$ can be written in the form $x,y,z,q^i(x,y,z)$.
Then the following relations are fulfilled:
$$
\frac{dq^i}{ds}=\dot q^i=q^i_{x}\dot x+q^i_{y}\dot y+
q^i_{z}\dot z
$$
and
$$
\frac{d^2 q^i}{ds^2}= \ddot q=q^i_{xx}(\dot x)^2+
2q^i_{xy}\dot x\dot y + q^i_{yy}(\dot y)^2+
2q^i_{xz}\dot x\dot z+ 2q^i_{yz}\dot y\dot z+
q^i_{zz}(\dot z)^2+q^i_x\ddot x+
q^i_y\ddot y+q^i_z \ddot z.
$$

     Putting these relations in the general equations of
geodesics:
$$
\frac{d^2 x^i}{ds^2}+\Gamma^i_{kj}\frac{d x^k}{ds}\frac{d x^j}{ds}=0.
$$
we obtain the system of equations for the functions $q^i(x,y)$
(after  making equal to zero the
expressions at the derivatives
$(\frac{d x}{ds})^2$,
$(\frac{dy}{ds})^2$,
$(\frac{d z}{ds})^2$,
and
$(\frac{d x}{ds}\frac{dy}{ds})$, $(\frac{d x}{ds}\frac{dz}{ds})$,
$(\frac{d y}{ds}\frac{dz}{ds})$.

     It has the form:
$$
q^i_{xx}=[\Gamma^1_{11}+2\Gamma^1_{1j}q^j_x+\Gamma^1_{kj}q^k_x q^j_x] q^i_x+
$$
$$
+[\Gamma^2_{11}+2\Gamma^2_{1j}q^j_x+\Gamma^2_{kj}q^k_x q^j_x] q^i_y +
$$
$$
+[\Gamma^3_{11}+2\Gamma^3_{1j}q^j_x+\Gamma^3_{kj}q^k_x q^j_x] q^i_z-
2\Gamma^i_{1j}q^j_x-\Gamma^i_{kj}q^k_x q^j_x-\Gamma^i_{11},
$$
$$
q^i_{xy}=[\Gamma^1_{12}+\Gamma^1_{1j}q^j_y+\Gamma^1_{2j}q^j_x+
\frac{1}{2}\Gamma^1_{kj}(q^k_x q^j_y+q^k_y q^j_x)] q^i_x +
$$
$$
+[\Gamma^2_{12}+\Gamma^2_{1j}q^j_y+\Gamma^2_{2j} q^j_x+
\frac{1}{2}\Gamma^2_{kj}(q^k_x q^j_y+q^k_y q^j_x)] q^i_y+
$$
$$
+[\Gamma^3_{12}+\Gamma^3_{1j}q^j_y+\Gamma^3_{2j} q^j_x+
\frac{1}{2}\Gamma^3_{kj}(q^k_x q^j_y+q^k_y q^j_x)] q^i_z-
$$
$$
-\Gamma^i_{1j}q^j_y - \Gamma^i_{2j} q^j_x-
\frac{1}{2}\Gamma^i_{kj}(q^k_x q^j_y+q^k_y q^j_x)-\Gamma^i_{12},
$$
$$
q^i_{xz}=[\Gamma^1_{13}+\Gamma^1_{1j}q^j_z+\Gamma^1_{3j}q^j_x+
\frac{1}{2}\Gamma^1_{kj}(q^k_x q^j_z+q^k_z q^j_x)] q^i_x +
$$
$$
+[\Gamma^2_{13}+\Gamma^2_{1j}q^j_z+\Gamma^2_{3j} q^j_x+
\frac{1}{2}\Gamma^2_{kj}(q^k_x q^j_z+q^k_z q^j_x)] q^i_y+
$$
$$
+[\Gamma^3_{13}+\Gamma^3_{1j}q^j_z+\Gamma^3_{2j} q^j_x+
\frac{1}{2}\Gamma^3_{kj}(q^k_x q^j_z+q^k_z q^j_x)] q^i_z-
\Gamma^i_{1j}q^j_z - \Gamma^i_{3j} q^j_x-
\frac{1}{2}\Gamma^i_{kj}(q^k_x q^j_y+q^k_y q^j_x)-\Gamma^i_{13},
$$
$$
q^i_{yz}=[\Gamma^1_{23}+\Gamma^1_{2j}q^j_z+\Gamma^1_{3j}q^j_y+
\frac{1}{2}\Gamma^1_{kj}(q^k_y q^j_z+q^k_z q^j_y)] q^i_x +
$$
$$
+[\Gamma^2_{23}+\Gamma^2_{2j}q^j_z+\Gamma^2_{3j} q^j_y+
\frac{1}{2}\Gamma^2_{kj}(q^k_y q^j_z+q^k_z q^j_y)] q^i_y+
$$
$$
+[\Gamma^3_{23}+\Gamma^3_{2j}q^j_z+\Gamma^3_{3j} q^j_y+
\frac{1}{2}\Gamma^3_{kj}(q^k_y q^j_z+q^k_z q^j_y)] q^i_z-
\Gamma^i_{2j}q^j_z - \Gamma^i_{3j} q^j_y-
\frac{1}{2}\Gamma^i_{kj}(q^k_z q^j_y+q^k_y q^j_z)-\Gamma^i_{23},
$$
$$
q^i_{yy}=[\Gamma^1_{22}+2\Gamma^1_{2j}q^j_y+\Gamma^1_{kj}q^k_y q^j_y] q^i_x+
[\Gamma^2_{22}+2\Gamma^2_{2j}q^j_y+\Gamma^2_{kj}q^k_y q^j_y] q^i_y +
$$
$$
+[\Gamma^3_{22}+2\Gamma^3_{2j}q^j_y+\Gamma^3_{kj}q^k_y q^j_y] q^i_z-
2\Gamma^i_{2j}q^j_x-\Gamma^i_{kj}q^k_y q^j_y-\Gamma^i_{22},
$$
$$
q^i_{zz}=[\Gamma^1_{33}+2\Gamma^1_{1j}q^j_z+\Gamma^1_{kj}q^k_z q^j_z] q^i_x+
[\Gamma^2_{33}+2\Gamma^2_{3j}q^j_z+\Gamma^2_{kj}q^k_z q^j_z] q^i_y +
$$
$$
+[\Gamma^3_{33}+2\Gamma^3_{1j}q^j_z+\Gamma^3_{kj}q^k_z q^j_z] q^i_z-
2\Gamma^i_{3j}q^j_z-\Gamma^i_{kj}q^k_z q^j_z-\Gamma^i_{33}.
$$

     So, in linear approximation we get the matrix Lame system of equations
on coordinates $q^i(x,y,z)$.

\section{ Equation of geodesic deviation and the matrix Darboux-Lame system}

     The general equation of geodesic deviation is
$$
\frac{D^2 \eta^i}{ds^2}+R^i_{kjm}\frac{dx^k}{ds}\frac{dx^m}{ds}\eta^j=0,
$$
where $R^i_{kjm}$ is the tensor of curvature of manifold.

    We transform this equation into an easy-to-use form. According to
 the rule of covariant differentiation we obtain the relation
$$
\frac{D \eta^i}{ds}=\frac{d\eta^i}{ds}+\Gamma^{i}_{jk}\eta^j \frac{dx^k}{ds}=
\rho^i
$$
and then
$$
\frac{D^2 \eta^i}{ds^2}=\frac{D \rho^i}{ds}=\frac{d\rho^i}{ds}+\Gamma^{i}_{lm}
\rho^l\frac{dx^m}{ds}=
$$
$$
=\frac{d^2 \eta^i}{ds^2}+\frac{\partial \Gamma^{i}_{jk}}{\partial x^l}\eta^j
\frac{dx^k}{ds}\frac{dx^l}{ds}+\Gamma^{i}_{jk}\frac{d \eta^j}{ds}\frac{dx^k}{ds}+
\Gamma^{i}_{jk}\eta^j\frac{d^2 x^k}{ds^2}+\Gamma^{i}_{lm}(\frac{d\eta^l}{ds}+
\Gamma^{l}_{jk}\eta^j\frac{dx^k}{ds})\frac{dx^m}{ds}.
$$

     Furthermore, using the equations of geodesics
$$
\frac{d^2 x^k}{ds^2}+\Gamma^{k}_{pq}\frac{dx^p}{ds}\frac{dx^q}{ds}=0,
$$
we can get the relation:
$$
\frac{d^2 \eta^i}{ds^2}+2\Gamma^{i}_{lm}\frac{dx^m}{ds}\frac{d\eta^l}{ds}+
\left[\frac{\partial \Gamma^{i}_{jk}}{\partial x^l}
\frac{dx^k}{ds}\frac{dx^l}{ds}+\right.
$$
$$
\left.+\Gamma^{i}_{lm}\Gamma^{l}_{jk}\frac{dx^k}{ds}
\frac{dx^m}{ds}-\Gamma^{i}_{jk}\Gamma^{k}_{pq}\frac{dx^p}{ds}\frac{dx^q}{ds}+
R^i_{kjm}\frac{dx^k}{ds}\frac{dx^m}{ds}\right]\eta^j=0.
$$

    Putting here the expression for the curvature tensor
$$
R^i_{kjm}=\frac{\partial \Gamma^{i}_{km}}{\partial x^j}-
\frac{\partial \Gamma^{i}_{kj}}{\partial x^m}+\Gamma^{i}_{nj}\Gamma^{n}_{km}-
\Gamma^{i}_{nm}\Gamma^{n}_{kj},
$$
we find the relation
$$
\frac{d^2 \eta^i}{ds^2}+2\Gamma^{i}_{lm}\frac{dx^m}{ds}\frac{d\eta^l}{ds}+
\left[\frac{\partial \Gamma^{i}_{jk}}{\partial x^l}
+\Gamma^{i}_{nl}\Gamma^{n}_{jk}-
\Gamma^{i}_{jn}\Gamma^{n}_{kl}+\right.
$$
$$
\left.+\frac{\partial \Gamma^{i}_{kl}}{\partial x^j}-
\frac{\partial \Gamma^{i}_{kj}}{\partial x^l}+\Gamma^{i}_{nj}\Gamma^{n}_{kl}-
\Gamma^{i}_{nl}\Gamma^{n}_{kj}\right]\frac{dx^k}{ds}\frac{dx^l}{ds}\eta^j,
$$
whence after rearragement the desired formula
for the equation of geodesic deviation follows
$$
\frac{d^2 \eta^i}{ds^2}+2\Gamma^{i}_{lm}\frac{dx^m}{ds}\frac{d\eta^l}{ds}+
\frac{\partial \Gamma^{i}_{kl}}{\partial x^j}\frac{dx^k}{ds}
\frac{dx^l}{ds}\eta^j=0.
\eqno (10)
$$

     This form of equation of geodesic deviation is convenient in
applications (see [14]). Let us consider some of them.

     In three-dimensional case one obtains
$$
\frac{d\eta^i}{ds}=\dot \eta^i=\eta^i_{x}\dot x+\eta^i_{y}\dot y+
\eta^i_{z}\dot z
$$
and
$$
\frac{d^2 \eta^i}{ds^2}= \ddot \eta^i=\eta^i_{xx}(\dot x)^2+
2\eta^i_{xy}\dot x\dot y + \eta^i_{yy}(\dot y)^2+
2\eta^i_{xz}\dot x\dot z+ 2\eta^i_{yz}\dot y\dot z+
\eta^i_{zz}(\dot z)^2+\eta^i_x\ddot x+
\eta^i_y\ddot y+\eta^i_z \ddot z.
$$
     Putting these relations in (10) and using the equations of
geodesics  we get the matrix Lame system of equations
$$
\eta^i_{xx}+2\Gamma^i_{1l}\eta^l_x -\Gamma^1_{11}\eta^i_x-
\Gamma^2_{11}\eta^i_y-\Gamma^3_{11}\eta^i_z+
\frac{\partial\Gamma^i_{11}}{\partial x^j}\eta^j=0,
$$
$$
\eta^i_{yy}+2\Gamma^i_{2l}\eta^l_y -\Gamma^1_{22}\eta^i_x-
\Gamma^2_{22}\eta^i_y-\Gamma^3_{22}\eta^i_z+
\frac{\partial\Gamma^i_{22}}{\partial x^j}\eta^j=0,
$$
$$
\eta^i_{zz}+2\Gamma^i_{3l}\eta^l_z -\Gamma^1_{33}\eta^i_x-
\Gamma^2_{33}\eta^i_y-\Gamma^3_{33}\eta^i_z+
\frac{\partial\Gamma^i_{33}}{\partial x^j}\eta^j=0,
$$
$$
\eta^i_{xy}+\Gamma^i_{11}\eta^1_y+\Gamma^i_{12}(\eta^2_y+\eta^1_x)+
\Gamma^i_{13}\eta^3_y+\Gamma^i_{23}\eta^3_x+\Gamma^i_{22}\eta^2_x-
\Gamma^1_{12}\eta^i_x-\Gamma^2_{12}\eta^i_y - \Gamma^3_{12}\eta^i_z+
\frac{\partial\Gamma^i_{12}}{\partial x^j}\eta^j=0,
$$
$$
\eta^i_{xz}+\Gamma^i_{11}\eta^1_z+\Gamma^i_{13}(\eta^3_z+\eta^1_x)+
\Gamma^i_{12}\eta^2_z+\Gamma^i_{23}\eta^2_x+\Gamma^i_{33}\eta^3_x-
\Gamma^1_{13}\eta^i_x-\Gamma^2_{13}\eta^i_y - \Gamma^3_{13}\eta^i_z+
\frac{\partial\Gamma^i_{13}}{\partial x^j}\eta^j=0,
$$
$$
\eta^i_{yz}+\Gamma^i_{12}\eta^1_z+\Gamma^i_{23}(\eta^3_z+\eta^2_y)+
\Gamma^i_{13}\eta^1_z+\Gamma^i_{33}\eta^3_z+\Gamma^i_{22}\eta^2_z-
\Gamma^1_{23}\eta^i_x-\Gamma^2_{23}\eta^i_y - \Gamma^3_{23}\eta^i_z+
\frac{\partial\Gamma^i_{23}}{\partial x^j}\eta^j=0.
$$

\section{Acknowledgement}

     This work was supported by INTAS-93-0166 and the author is grateful to
the Physics Departement of Lecce University for the financial support and
hospitality.
\bigskip

\section{References}
\begin{itemize}
\item[1] {Zakharov V., Manakov S.V.: Construction of Multidimensional
Nonlinear Integrable system and their Solitons,
Funk. Anal.i ego prilozh., v.19,  N0 2, p. 11--25, 1985}
\item[2]{ Dryuma V.S.: Proektivnye svo\u istva kommutiruyushchih seme\u istv
operatorov, $IX$ Vsesoyuznaya geometricheskaya konferentsia:
Tezisy  dokladov, 1988, 20--22 sentyabrya, Kishinev: \c Stiin\c te, s. 104--105}
\item[3]{Dryuma V.S.: 3-dim exactly integrable system of nonlinear equations,
 Workshop NEEDS'90, talk, Dubna, 1980}
\item[4]{Dryuma V.S.: 4-dim exactly integrable system of nonlinear equations,
Rep. of IX International conference on Topology and Applications, Kiev, 12--16 
October, 1992, p. 71}
\item[5]{Dryuma V.S.: On geometry of the second order nonlinear differential 
equations, Proc. of Conference "Nonlinear Phenomena", M.: Nauka, 1992, 
p. 41--48}
\item[6]{Three dimensional exactly integrable system of equations and its 
applications, Mathematical studies, 1994, v.124, N.3, p.56--68}
\item[7]{Sobchuk V.: Ob odnom klasse normalnyh rimanovyh prostranstv,
Trudy seminara po vektornomu i tenzornomu analizu, t. $XV$.
             M.: MGU, 1970, s. 65--78}
\item[8]{ Konopelchenko B.G.: Solitons in multidimensions, 1993, 
World Scientific, Singapor}
\item[9]{ Krichever I.: Algebraic-geometrical n-orthogonal curvilinear
coordinate systems and solutions to the associativity
equations. Preprint, 1996, vol. 1, no. 1, p. 1--25}

\item[10]{ Zakharov V.E.: Description of the n-orthogonal curvilinear coordinate
             systems and Hamiltonian Integrable Systems of Hydrodynamic Type:
             Part 1. Integration of the Lame equations.
 Preprint, 1997, vol. 1, no. 1, p. 1--42}

\item[11]{ Holzapfel R.P.: Euler partial Differential Equations, 1986.
VEB Deutscher Verlag der Wi\-ssen\-schaf\-ten,
Berlin}
\item[12]{Dryuma V.S.: Geometrical applications of multidimensional integrable equations
Theor. Math. Phys., 1994, vol. 99, no. 3, p. 241--252}
\item[13]{Dryuma V.S.: Geometrical properties of nonlinear dynamical 
systems with regular and chaotic behaviour,
Proc. of the First Workshop on "Nonlinear Physics",
             World Scientific, Singapore, 1995, p. 83--93}

\item[14]{ Dryuma V.S., Konopelchenko B.G.: On equation of geodesic 
deviation and its solutions. Buletinul Academiei de \c Stiin\c te a 
Republicii Moldova,
             Matematica, 1996, no. 3(22), p. 61--73}
\end{itemize}

\enddocument